\documentclass[
aip, jap
 amsmath,amssymb,
 reprint,%
]{revtex4-1}
\usepackage{amsmath}  % in your preamble
\usepackage{graphicx}% Include figure files
\usepackage{dcolumn}% Align table columns on decimal point
\usepackage{bm}% bold math
\usepackage{array} % for advanced table column definitions
\usepackage{nomencl}
\makenomenclature
\usepackage[utf8]{inputenc}
\usepackage[T1]{fontenc}
\usepackage{mathptmx}
\usepackage{etoolbox}
\usepackage{etoolbox}
\usepackage{textcomp}
\usepackage{gensymb}
\usepackage{textcomp}
\usepackage{siunitx}

\makeatletter
\def\@email#1#2{%
 \endgroup
 \patchcmd{\titleblock@produce}
  {\frontmatter@RRAPformat}
  {\frontmatter@RRAPformat{\produce@RRAP{*#1\href{mailto:#2}{#2}}}\frontmatter@RRAPformat}
  {}{}
}%
\makeatother
\begin{document}

\preprint{AIP/123-QED}

\title[Article]{High-throughput Parasitic-independent  Probe Thermal Resistance Calibration for Robust Thermal Mapping with Scanning Thermal Microscopy}
% Force line breaks with \\
\author{Ram Munde}
\author{Heng-Ray Chuang}
\affiliation{Department of Materials Engineering, Purdue University, West Lafayette, USA}
% \author{Chongke Gu}
% \affiliation{Elmore Family School of Electrical and Computer Engineering, Purdue University, West Lafayette, IN USA}

\author{Raisul Islam$^*$}
\email{raisul@purdue.edu}
\affiliation{Department of Materials Engineering, Purdue University, West Lafayette, USA}
\affiliation{Elmore Family School of Electrical and Computer Engineering, Purdue University, West Lafayette, IN USA}

 %\homepage{https://engineering.purdue.edu/RISE-Lab.}
% \affiliation{%
% Second institution and/or address%\\This line break forced% with \\
% }%

\date{\today}% It is always \today, today,

\begin{abstract}
Nanostructured materials, critical for thermal management in semiconductor devices, exhibit a strong size dependence in thermal transport. Studying thermal resistance variation across grain boundaries is critical for designing effective thermal interface materials. Frequency-domain Thermoreflectance (FDTR)-based techniques can provide thermal resistance mapping at the micrometer ($\mu$m) scale. Scanning Thermal Microscopy (SThM) enables quantification of local thermal transport with significantly higher spatial resolution ($<100$~nm). However, challenges in quantifying the raw signal to thermal conductivity and surface sensitivity limit its widespread adoption for understanding nanoscale heat transport and defect-mediated thermal properties in nanostructured films. Here, we introduce a circuit-based probe thermal resistance ($R_p$) calibration technique independent of parasitic heat pathways, enabling accurate determination of probe heat dissipation and tip temperature rise, thereby allowing extraction of local thermal resistance. SThM achieved sub-100~nm spatial resolution in mapping thermal resistance across a 15~nm-thick Al film deposited via e-beam evaporation on SiO$_2$ substrate. The thermal resistance maps are converted to thermal conductivity using robust analytical and finite element models that account for tip-sample geometry, lateral heat spreading, and buried interface effects. Gaussian distribution fitting of pixel-level thermal resistance values yields $k_{\mathrm{Al}} = 45.1^{+4.7}_{-3.6}$~W/(m$\cdot$K) for the ultra-thin Al film (13--15~nm), representing a 5.3-fold reduction from bulk aluminum (237~W/(m$\cdot$K)). These results agree with published experimental data and theoretical frameworks explaining thickness-dependent heat transport in ultra-thin metallic films.

\end{abstract}

\maketitle

\printnomenclature
\hrule

\nomenclature{$\tau_e$}{Electronic relaxation time ($\mathrm{s}$)}
\nomenclature{$\lambda_e$}{Electron Mean Free Path}
\nomenclature{$\tau$}{Average scattering time ($\mathrm{s}$)}
\nomenclature{$\lambda$}{Wavelength ($\mathrm{m}$)}
\nomenclature{$NA$}{Numerical aperture}
\nomenclature{$\mu$}{Thermal diffusion length ($\mathrm{m}$)}
\nomenclature{$\alpha$}{Thermal diffusivity ($m^2/s$)}
\nomenclature{$f$}{Frequency ($1/s$)}
\nomenclature{b}{Probe tip radius}
\nomenclature{\Omega}{Tip electrical resistance}
\nomenclature{\Omega$_p$}{Probe electrical resistance}
\nomenclature{$R_{ss}$}{Solid-solid thermal resistance}

\section{\label{sec:level1} Introduction}
\begin{figure*}[htbp]
\includegraphics{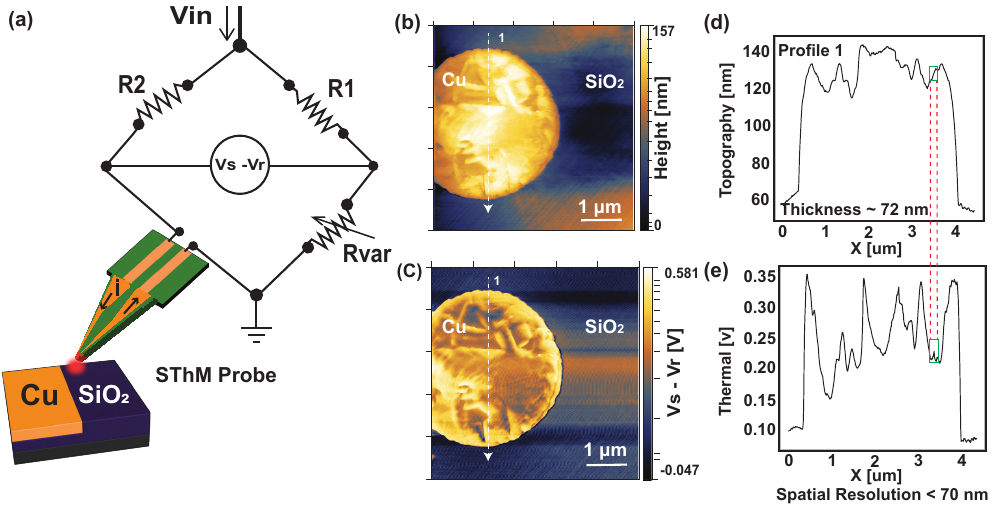}% Here is how to import EPS art
\caption{\label{fig:intro} (a) Schematic of a Kelvin Nanotechnology (KNT) thermistor probe with an integrated Wheatstone bridge for high-precision thermal signal detection. The input voltage ($V_{in}$) supplied using integrated software, while variable resistor balances the bridge by matching tip resistance at the equilibrium conditions. (b) SThM topography and (c) SThM thermal signal maps of a 5 $\mu$m × 5 $\mu$m Cu-Through Silicon Via (TSV) sample. The thermal signal in (c) reveals local thermal conductivity contrast arising from grain boundary and surface roughness variation. (d) SThM topography [nm] signals and (e) SThM thermal signal [V] profiles along line 1. Consecutive changes in thermal signal peaks, with no corresponding change in topography, demonstrate that the spatial resolution of local thermal conductivity mapping is < 70\,nm.}
\end{figure*}
Across the last two decades, continuous materials scaling coupled with precise nanostructuring has powered a remarkable run of progress in electronic\cite{10019538,munde20253d, banerjee20023}, photonic\cite{ostroverkhova2016organic, huang2020recent}, and quantum technologies\cite{putranto2024deep}, reinforced by advances in MEMS/NEMS \cite{zhu2019development, lyshevski2018mems}, nanoscale sensors for biomedicine\cite{hannah2020developments, wang2013advances, tovar2023recent}, and scanning-probe/optical instrumentation\cite{cahill2003nanoscale}. Miniaturization fundamentally modifies transport, mechanical, and interfacial behavior. As a result, size-dependent variations in material properties become critical considerations in modeling, metrology, and device design. For example, silicon (Si) thermal conductivity ($k$) \nomenclature{$k$}{Thermal conductivity (W/m$\cdot$K)} is reduced from its bulk $k$  value of 138 W/m$\cdot$K to 2.2 W/m$\cdot$K  for a 155\,nm Si thin film\cite{mcconnell2005thermal, huang2019thermal}. When the physical dimension becomes comparable to or smaller than the mean free path \nomenclature{$\Lambda$}{Mean free path (m)} ($\Lambda$) of heat carriers (electrons and phonons), the macroscopic equilibrium and continuum assumptions break down. Therefore, understanding the nanoscale energy transport phenomena, both analytically and experimentally, is crucial for the continued miniaturization of technology.

Thermal metrology plays a vital role in validating theoretical models and advancing the understanding of energy transport mechanisms. Conventional techniques such as $\mu$-thermocouples, microbridge and suspended devices ($3\omega$), infrared thermography (IR), and liquid crystal thermography are used to measure macroscopic thermal properties of devices. Recently, with technological advancements in device fabrication and deposition processes, the $3\omega$ technique has been extended for use with thin films and microscale devices by fabricating bridges a few $\mu m$ thick  \cite{cahill1987thermal}. This method enables accurate characterization of both in-plane and cross-plane thermal transport in metallic, semiconducting, and dielectric thin films. A major limitation of the 3$\omega$ technique is its requirement for complex sample preparation and the complexities arising from voltage oscillations when measuring metals and semiconductors, thus limiting its applicability to a narrow range of materials\cite{cahill2002thermometry}. Optical based techniques such as frequency- and time-domain thermoreflectance spectroscopy (FDTR and TDTR), Raman spectroscopy, laser flash techniques, and fluorescence techniques are widely used for measuring the temperature and thermal transport properties of microstructured materials \cite{christofferson2008microscale, tiwari2013anisotropic, zeng2017photothermal}. In contrast to the $3\omega$ method, optical-based techniques such as TDTR and FDTR enable non-contact characterization of thermal properties, minimizing physical interference with the sample. Recent advancements in laser source technology have significantly enhanced the capability of TDTR and FDTR techniques to characterize the thermal properties of materials with superior temporal and spatial resolution \cite{isotta2023microscale}. Depending on the type of laser used in the measurement, these methods can achieve temporal resolutions ranging from a few nanoseconds to $\sim$ 30 fs (1 fs = $10^{-15}$ sec). It is worth noting that the electronic relaxation time ($\tau_e$) typically lies in the range of about $\sim$10–50 fs. Therefore, TDTR and FDTR are useful techniques for measuring the thermal transport properties associated with electron-phonon coupling mechanisms. However, these optical-based thermal measurement methods suffer from an intrinsic spatial resolution limitation imposed by the pump–probe spot size (typically $\sim$0.5–10 $\mu$m), rendering them inadequate for reliably imaging thermal heterogeneities arising from nanoscale defects.

In the late 1990s, the development of scanning probe microscopy (SPM)–based techniques, such as scanning thermal microscopy (SThM) and near-field scanning optical microscopy (NSOM) provided unprecedented spatial resolution for nanoscale characterization and manipulation. In general, NSOM requires optically compatible and flat samples, is limited by weak signal-to-noise ratio, and exhibits greater tip fragility compared to SThM. Table \ref{tab:thermal_methods} summarizes the quantitative capabilities of thermal metrology techniques for microelectronics. Current commercially available SThMs are based on the working principle designed by Majumdar \emph{et al.}\cite{majumdar1993thermal} as illustrated in Fig. \ref{fig:intro}(a). The SThM system uses specialized thermal probes (TPs) with tip radii below 100\,nm, capable of resolving temperature differences smaller than 0.1 $^\circ C$ with sub-100\,nm spatial resolution, thereby enabling quantitative nanoscale thermal characterization of materials. SThM can operate in either temperature contrast mode (TCM) or conductivity contrast mode (CCM). In TCM, a tip-integrated thermocouple measures the junction temperature at the tip-sample interface, whereas the CCM mode provides qualitative mapping of local thermal conductivity across a broad range of materials, including metallic, semiconducting, and dielectric thin films. Transforming these qualitative signals into quantitative thermal conductivity values is not trivial because of the complex coupled thermal transport at the tip-sample interface. There is no standardized technique for obtaining quantitative measurements of thermal resistance with SThM. Recently, various theoretical modeling, calibration strategies, and instrument add-on techniques have been reported\cite{gucmann2021scanning,li2022comprehensive,li2025quantitative}. Bodzenta \emph{et al.} and Li \emph{et al.} provided comprehensive reviews of SThM theoretical models and experimental strategies for quantitative thermal measurements\cite{bodzenta2022scanning, li2025quantitative}.

In this work, we present a circuit-based calibration technique for measuring non-contact thermal resistance ($R_{nc}$), also known as probe thermal resistance. Compared with conventional thermal-stage methods, this calibration approach reduces sensitivity to temperature-controller errors and environmental factors (e.g., moisture), and is less affected by common artifacts, including calibration-sample variations. First, we construct thermal-resistance maps across the film surface by coupling scanning thermal measurements with finite-element heat-transfer modeling of the tip–sample mechanical contact. These maps resolve spatial variations in thermal transport that arise from differences in crystallite size and orientation, effects that are most pronounced in polycrystalline and amorphous films. Because the maps carry the imprint of surface preparation and condition—the same steps a film undergoes in routine device-integration flows—the resulting resistance estimates reflect the as-processed device film. Building on the same calibrated contact model, we then extract the film's effective thermal conductivity ($k_{eff}$) using an experimentally anchored procedure that combines accurate instrument calibration with contact heat-transfer modeling. Sec. \ref{sec:level2} of this paper discusses the theoretical modeling of SThM required for understanding different heat transfer mechanisms at the tip-sample contact. Sec. \ref{sec:methodology} describes the methodology used to extract $k_{eff}$ of samples and the post-processing steps to study the underlying nanoscale thermal transport physics and to map the corresponding thermal resistances. In Sec. \ref{sec:results}, we present the results obtained through the techniques described in Sec.s \ref{sec:level2} and \ref{sec:methodology}. Finally, we summarize our results with concluding remarks.

\begin{table*}[t]
\caption{\textbf{ { Thermal Metrology Techniques for Microelectronics } }}
\label{tab:thermal_methods}

\begin{ruledtabular}
\begin{tabular}{l l c c}
Technique &
Mechanism &
Spatial Resolution &
Temperature Resolution \\
\hline

Infrared thermography &
Thermal radiation (Planck, Stefan--Boltzmann) &
3--10\,$\mu$m &
0.02--1 K \\

Photothermal Raman &
Raman shift or Stokes/anti-Stokes ratio &
0.5--1\,$\mu$m &
1--10 K \\

3$\omega$ method &
AC Joule heating with electrical sensing &
--- &
0.1 mK--1 K \\

TDTR / FDTR &
Reflectivity-based transient thermometry &
1--10\,$\mu$m &
$\sim$0.01 K \\

NSOM &
Near-field optical heat detection &
50--100\,nm &
0.1--1 K \\

\textbf{SThM} &
\textbf{AFM-integrated thermocouple/Pd thermistor tip} &
\textbf{10--50\,nm} &
\textbf{$\sim$0.1 K} \\

\end{tabular}
\end{ruledtabular}

\vspace{2pt}
\footnotesize Data adapted from Refs.~\cite{christofferson2008microscale,cahill1987thermal,zhang2007nano}.
\end{table*}

\begin{figure}
\includegraphics[width=0.5\textwidth]{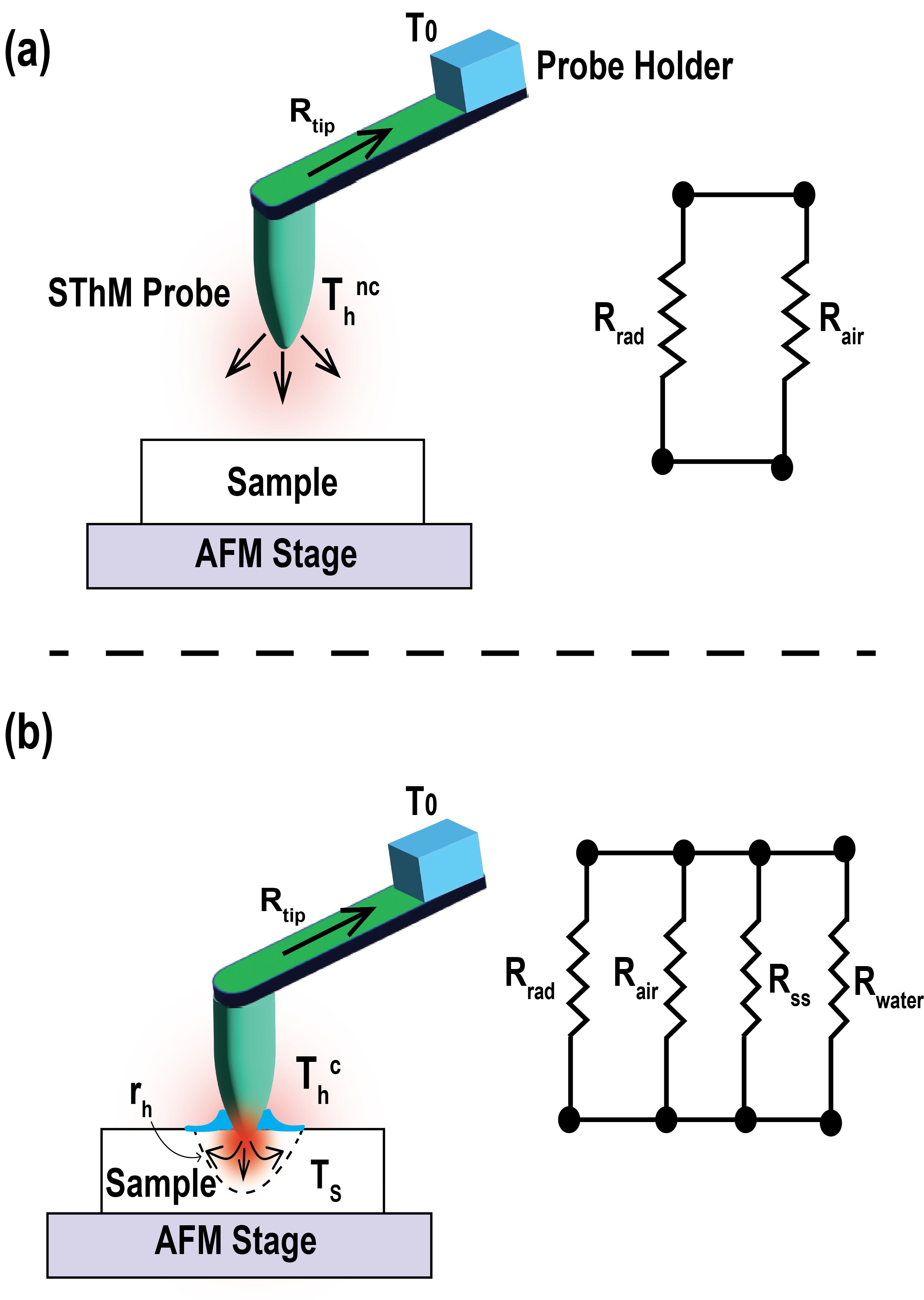}% Here is how to import EPS art
\caption{\label{fig:channel} {(a) Heat transfer pathways of the probe just before sample contact. The thermal resistances shown include: $R_{tip}$ (heat dissipation at the tip), $R_{rad}$ (radiative thermal resistance), and $R_{air}$ (convective thermal resistance). (b) Heat transfer pathways when the probe is in contact with the sample, introducing additional thermal resistance channels through the solid-solid ($R_{ss}$) contact which includes $R_{tip}$ and $R_{water}$ is the conduction due to water meniscus forming at tip-sample contact.}}
\end{figure}

\section{\label{sec:level2} Background of The Quantitative SThM Measurement Technique}
A schematic of the SThM probe with an integrated Wheatstone bridge is shown in Fig.~\ref{fig:intro}(a). The Wheatstone bridge helps detect changes in the thermal resistance of the temperature-sensitive tip material. SThM provides qualitative maps of both the thermal signal and the surface topography of the scanned region, as shown in Fig.~\ref{fig:intro}(b,c). The topography map reveals the nanoscale geometrical features of the surface, whereas the thermal signal maps offer qualitative insights into the local thermal conductivity of the sample with nanoscale spatial resolution. Furthermore, SThM exhibits a high signal-to-noise ratio, enabling more accurate and reliable measurements (see Fig.~\ref{fig:intro} (d,e)). However, correlating these measured signals with the intrinsic thermal properties of the materials requires a thorough theoretical understanding of the underlying physical mechanisms.

Theoretical modeling links the probe voltage signals measured by the Wheatstone bridge detector to the sample’s thermal properties. Therefore, accurate quantitative measurements of thermal conductivity heavily rely on mathematical modeling to transform probe voltage signals into thermal conductivity values. Thermal transport between the probe and sample involves multiple heat transfer channels, including conduction through the solid–solid contact (tip–sample) and the water meniscus (from the moisture), convection through air, and thermal radiation, which becomes significant at nanometer scale tip–sample distances (see Fig. \ref{fig:channel}). Each of these channels has its corresponding contribution, which depends on the instrumentation, measurement environment, and sample under consideration. Various analytical and numerical approaches have been reported for studying each heat transfer channel and its significance in quantitative measurements. A recent review by Li \emph{et al.} comprehensively discusses multiple analytical and numerical approaches for calculating heat transfer through these channels \cite{li2025quantitative}.

This work uses a thermistor probe tip that consists of Palladium (Pd) resistor material with a thermal coefficient of resistance (TCR) of 3800 $ppm/^\circ\mathrm{C}$, as specified by the manufacturer. Importantly, the TCR value is not affected by the experimental procedure or daily use of the probe\cite{pernot2021frequency, huang2024violation}. Since Pd is a metal, within a relatively small temperature range, the electrical resistance of the tip ($\Omega$) varies linearly with temperature (T).
\begin{equation}
\Omega = \Omega_o \left[1 + \alpha (\widetilde T - T_0)\right]
\label{eq:resistance}
\end{equation}
where $\Omega_o$ is the electrical resistance of the tip at temperature $T_0$ and $\alpha$ is the TCR. The tip temperature is measured by applying an electric current to the probe and monitoring the change in electrical resistance of the tip with SThM software.

When the probe is in air without contacting the sample, heat is dissipated by conduction through the cantilever, convection through the air, and radiation to the ambient, as shown in Fig.~\ref{fig:channel}(a). The heat transport is governed by Fourier’s law of conduction, as presented in Eq. (\ref{eq:heatnc}). The steep tip–sample temperature gradient may challenge the applicability of Fourier’s law. However, when the mean free path ($\Lambda$) of the dominant heat carriers is much smaller than the tip radius ($\sim$50 nm), and the SThM operates at slow scan rates (0.5–1 Hz, corresponding to $>1$ ms per pixel), local thermal equilibrium is maintained.\cite{deshmukh2022direct,swoboda2023spatially}. Therefore, Fourier’s law with an effective thermal conductivity can still be applied as a simple analytical model of heat transport. When the tip contacts the sample, a solid–solid tip–sample junction introduces an additional heat-conduction pathway, as described by Eq.~(\ref{eq:heatc}). Based on the heat-transfer channels illustrated in Fig.~\ref{fig:channel}(a,b), the relationships between the thermal resistances in non-contact and contact configurations are given by Eqs.~(\ref{eq:rnc_thermal}) and~(\ref{eq:rc_thermal}), respectively.

\begin{equation}
Q_h = \frac{\widetilde T_h^{nc} - T_0}{R_{nc}}
\label{eq:heatnc}
\end{equation}

\begin{equation}
Q_h = \frac{\widetilde T_h^{c} - T_0}{R_{nc}} + \frac{ \widetilde T_h^{c}-T_s}{R_{c}}
\label{eq:heatc}
\end{equation}

\begin{equation}
    \frac{1}{R_{nc}} = \frac{1}{R_{rad}}+\frac{1}{R_{air}}+\frac{1}{R_{tip}}
    \label{eq:rnc_thermal}
\end{equation}

\begin{equation}
    \frac{1}{R_c} = \frac{1}{R_{rad}}+\frac{1}{R_{air}} + \frac{1}{R_{ss}}+\frac{1}{R_{water}}
    \label{eq:rc_thermal}
\end{equation}

\nomenclature{$Q_h$}{Current driven Joule heat}
\nomenclature{$\widetilde T_h^{nc}$}{Average noncontact tip temperature}
\nomenclature{$T_0$}{Ambient temperature}
\nomenclature{$R_{nc}$}{Total noncontact thermal resistance}
\nomenclature{$\widetilde T_h^{c}$}{Average contact tip temperature}
\nomenclature{$T_s$}{Temperature of the sample}
\nomenclature{$R_c$}{Total contact thermal resistance}
\nomenclature{$V^{nc}$}{Noncontact probe voltage}
\nomenclature{$V^c$}{Contact probe voltage}

The current applied to the probe through the integrated SThM power controller induces Joule heating in both the probe and the tip material. The probe current in CCM is higher than that in TCM. Under the constant-current condition of CCM, the generated electrical heat ($Q_h$) flows from the tip into the sample. Therefore, the amount of heat transfer depends on the thermal conductivity of the sample. For highly thermally conductive materials, heat dissipates rapidly, resulting in a small temperature rise at the tip and a correspondingly low output voltage. In contrast, materials with high thermal resistance impede heat flow, causing heat to accumulate near the tip and producing a larger temperature rise and higher voltage output. The corresponding temperature change ($\Delta T$) can be expressed as shown in Eq.~(\ref{eq:deltaT}). Experimental verification of this relation is provided in Fig. S3 (Supplementary Material).

\begin{equation}
\Delta T = a V\label{eq:deltaT}
\end{equation} 
Using Eq.~(\ref{eq:deltaT}), the thermal signals can be converted into the corresponding tip temperature as follows:

\begin{equation}
   \Delta T^{nc} = \widetilde T_h^{nc} - T_0 =a V^{nc} 
   \label{eq:deltaTnc}
\end{equation}
\begin{equation}
   \Delta T^c = \widetilde T_h^c - T_0 =a V^c 
   \label{eq:deltaTc}
\end{equation}

Therefore, by substituting Eqs. (\ref{eq:deltaTnc}) and (\ref{eq:deltaTc}) into Eqs. (\ref{eq:heatnc}) and (\ref{eq:heatc}), we can obtain the relationship between the probe voltage and probe thermal resistance, as shown in Eq. (\ref{eq:wideeq}):

\begin{widetext}
\begin{equation}
    \frac{\Delta T^{nc} - \Delta T^{c}}{\Delta T^{nc}}
    = \frac{V^{nc} - V^{c}}{V^{nc}}
    = \frac{\widetilde T_h^{nc} - \widetilde  T_h^c}{\widetilde T_h^{nc} - T_0}
    = \frac{1}{R_{nc} + R_{c}} \left( R_{nc} + \frac{T_0 - T_{s}}{Q_{h}} \right)
    \label{eq:wideeq}
\end{equation}
\end{widetext}

Furthermore, since the tip radius (b) is very small, the associated thermal radius ($r_h$) is negligible relative to the sample size (>$10~  {mm}^2$). Thus, under thermal equilibrium \cite{gu2002imaging}, it is reasonable to approximate $T_0 \approx T_s$. Consequently, we obtain the following expression:
\begin{equation}
   \frac{V^{nc} - V^{c}}{V^{nc}} = \frac{R_{nc}}{R_{nc}+R_c}
   \label{eq:voltagethermalresistance}
\end{equation}
Therefore, using Eq.~(\ref{eq:voltagethermalresistance}), the values of $R_c$ and $R_{nc}$ can be expressed in terms of the measured thermal signals ($V^c$ and $V^{nc}$). Here, $R_c$ embodies the sample’s thermal transport characteristics, and its detailed formulation will be discussed in Sec. \ref{sec:methodology}.

\section{\label{sec:methodology}Methods}
\subsection{Sample preparation}
A 5 mm × 10 mm Al film is patterned on a SiO$_2$/Si substrate. A 6-inch CZ-prime wafer with a 1.5 $\mu$m SiO$_2$ film on its surface was diced into 12 mm × 12 mm pieces. The p-type (B doped) resistivity of the substrate is between 10-25 $\Omega$-cm. The surface was first cleaned by sequentially sonicating in toluene, acetone, and 2-propanol for 5 minutes, then rinsed with DI water. During the process, non-polar and polar contaminants were cleaned with toluene and acetone, respectively, and 2-propanol was used to remove acetone from the substrate. A positive photoresist (AZ1518) was spin-coated at 3000 rpm for 60 s with a dwell time of 5 s, which was then followed by a soft bake at 110 $\celsius$ for 60 s. The substrate was exposed using a Heidelberg MLA150 maskless aligner, with the laser set to 405\,nm and a dose of 160 mJcm$^{-2}$. The sample was then developed in MF-26A for 35 s to remove the exposed photoresist over an area of 5 mm × 10 mm. A 15\,nm Al film was evaporated by CHA e-beam evaporator. The evaporation was controlled by a Telemark 861 deposition controller, and the deposition rate was 2 Ås$^{-1}$. A lift-off step was applied to remove the photoresist and Al from the unexposed area, leaving a 5 mm × 10 mm Al  film on the surface.  

\subsection{Experimental setup and numerical data generation}
The thermal signal data are obtained from the experiments conducted using an AFM (Dimension Icon\textsuperscript{\textregistered} manufactured by Bruker Inc.) equipped with an integrated SThM module. The SThM probe is purchased from Bruker Corp. and it consists of a thin Pd line on Si$_3$N$_4$ cantilever, as shown in Fig. S5 (Supplementary Material).  This probe is pre-mounted on the probe holder, and the probe tip serves as one of the arms of the Wheatstone bridge, as shown in the schematic in Fig.~\ref{fig:intro}(a). Before each SThM scan, the Wheatstone bridge is balanced to minimize background thermal signal noise. Subsequently, an input voltage (> 0.5 V) is applied to the probe. The temperature increase at the SThM tip changes the resistance of the Pd heater, producing a voltage imbalance across the Wheatstone bridge. This imbalance is detected as the SThM voltage signal ($V_s - V_r$). All measurements were conducted at a constant contact setpoint force and scan rate to ensure consistency. We used a gain of $1000\times$ while extracting the change in voltage of the Pd heater. This gain helps amplify the corresponding thermal signals without any loss of information.

For thermal relaxation and steady-state measurements, the SThM tip rasters at a relatively low scan rate of 0.5--1 Hz (S2 supplementary material). This scan rate is maintained throughout the experiment to reduce variance in thermal signals due to thermal drift. The acquired images are analyzed using Gwyddion software, while the pixel-wise thermal signal data are extracted and processed through a specially developed Python code tailored for this study.
\subsection{Measuring probe thermal resistance ($R_{nc}$)} % Here
Understanding the heat transfer between the probe and the surrounding system is essential for the quantitative measurement of thermal conductivity. The probe dissipates a finite amount of heat into the surrounding environment even in the non-contact state. Accurate quantification of this background signal is necessary to isolate the intrinsic thermal contribution from the sample. In earlier reports, the measurement of non-contact thermal resistance ($R_{nc}$) was carried out by balancing the tip temperature using a thermal stage \cite{huang2024violation, deshmukh2022direct}. This method, however, is prone to inaccuracies arising from parasitic thermal losses due to the roughness of the stage, non-uniform temperature profiles, and additional convective heat transfer to the probe.

\begin{figure}
\includegraphics{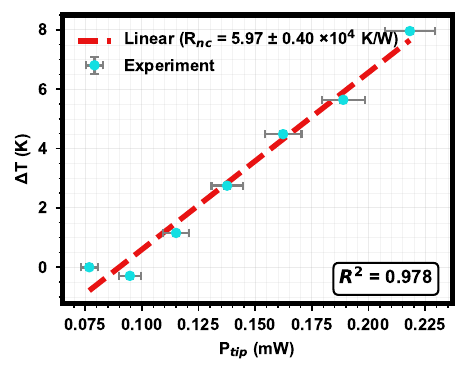}% Here is how to import EPS art
\caption{\label{fig:Rp} SThM non-contact thermal resistance ($R_{nc}$) measured with monitored Joule heating. }
\end{figure}

We propose a circuit-based technique to measure the non-contact thermal resistance ($R_{nc}$) by monitoring the probe-tip Joule heating power and the resulting temperature rise under an applied voltage. Unlike stage-based techniques, this approach eliminates dependence on thermal stage sensitivity and minimizes the influence of parasitic thermal effects from the environment. In this approach, Joule heating from the applied current raises the tip temperature. The temperature rise is directly proportional to the input power, where the proportionality constant corresponds to the non-contact thermal resistance, $R_{nc}$ (Eq.\ref{eq:rnc}). The temperature rise ($\Delta T$) was calculated relative to the initial tip temperature, which was at room temperature ($T_0  \sim 294 $  K), and the corresponding initial electrical resistance of the tip ($\Omega_o \sim$ 177 $\Omega$). It is important to note that the probe electrical resistance ($\Omega_p$), which is output of balanced Wheatstone bridge, differs from the tip resistance ($\Omega$) due to the presence of built-in limiting resistors in the probe, as shown in Fig. S4 (Supplementary Material). The applied power ($P_{tip}$) was determined directly from Joule heating in the tip under the applied current using Eq.(\ref{eq:power}). The resulting data were fitted using linear regression, yielding an $R_{nc}$ value of $(5.97 \pm 0.40) \times 10^4$ K/W as shown in Fig.~\ref{fig:Rp}. An instrumental uncertainty of $\pm$ 0.1 K in $\Delta T$ due to laser illumination was considered in the analysis\cite{spiece2019quantitative}. It should be noted that the uncertainty associated with laser illumination may be relatively large; however, since the analysis focuses on the relative change in temperature ($\Delta T$), the impact of this uncertainty is reduced. This result is consistent with previously reported values for similar probes, thereby validating the accuracy of the proposed measurement technique \cite{huang2024violation, razeghi2023single}.
\begin{equation}
    P_{tip}= \frac{V_{tip}^2}{\Omega(\widetilde T)}
    \label{eq:power}
\end{equation}
\begin{equation}
    R_{nc} = \frac{\Delta T}{P_{tip}}
    \label{eq:rnc}
\end{equation}
Here, $V_{tip}$ denotes the voltage drop across the tip, extracted from the known SThM probe resistance and the Wheatstone bridge input voltage (Fig.~\ref{fig:intro}(a)).

\begin{figure*}[htbp]
\includegraphics{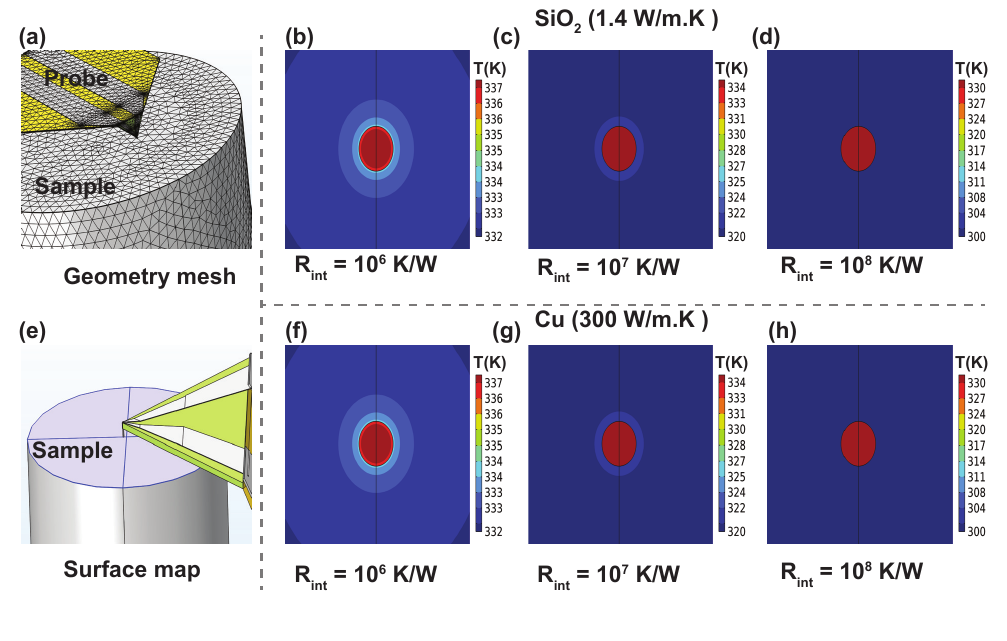}
\caption{\label{fig:rint_surface} (a) Custom meshed geometry of the SThM probe and the sample. The probe consists of a Si$_3$N$_4$ cantilever with Au connection pads and a Pd tip resistor. A 5\,nm Ni–Cr thin layer is deposited between the Au pads and the Pd tip to suppress reverse heat transfer and to avoid sudden increases in Joule heating. (b) Effect of $R_{int} = 10^6 \mathrm{K/W}$ on the surface temperature of the SiO$_2$ sample and the SThM tip contact. (c) Temperature variation at the contact interface for $R_{int} = 10^7 \mathrm{K/W}$. (d) Reduced surface heating due to the higher thermal resistance of $R_{int} = 10^8 \mathrm{K/W}$. (e) Schematic representation of the sample surface in contact with the SThM tip. (f–h) Effect of $R_{int}$ on the Cu surface for $10^6$, $10^7$, and $10^8 \mathrm{K/W}$, respectively.
}
\end{figure*}
\subsection{Modeling contact thermal resistance ($R_c$)}
As discussed in Sec. \ref{sec:level2},  $R_c$ is an experimentally measured quantity that comprises the thermal properties of the sample. In addition to solid-solid thermal resistance ($R_{ss}$), $R_c$ also includes thermal resistances arising from radiation ($R_{rad}$), air conduction ($R_{air}$) and the water meniscus formed at the tip–sample contact ($R_{water}$). The contribution of radiative heat transfer is often considered negligible, particularly under ambient SThM conditions with a cantilever–sample separation larger than 1 $\mu$m \cite{li2025quantitative}. The probe considered in our measurement has a tip height of $\sim$10 $\mu m$ measured from the cantilever. In a humid environment, as shown in Fig.~\ref{fig:channel}(b), a water meniscus forms at the tip–sample contact and serves as an additional pathway for heat transfer. A detailed study of water-meniscus formation and a theoretical model for heat transport through the meniscus were presented in Refs.~\onlinecite{gomes1999dc} and \onlinecite{luo1997sensor}, respectively. Furthermore, Refs.~\onlinecite{zhang2020review} and \onlinecite{assy2015temperature} reported that the contribution of the meniscus to the total tip–sample contact resistance ($R_c$) lies between 1$\%$ and 6$\%$. Therefore, the water-meniscus effect is treated as a 6$\%$ uncertainty in the $R_c$ values. Therefore, under the assumption that $R_{nc} \approx R_{tip}$, we express the measured quantity $(R_c - R_{nc})_{meas}$ as the sum of three thermal resistances, as shown in Eq.~(\ref{eq:seriesrth}).
\begin{equation}
    (R_c - R_{nc})_{meas} \approx R_{ss} = R_{\text{int}} + R_{\text{spr}}+ f \cdot R_{\text{BD}}
    \label{eq:seriesrth}
\end{equation}

where:
\begin{itemize}
    \item $R_{\text{int}}$: Tip-sample interface thermal resistance
    \item $R_{\text{spr}}$: Spreading resistance (Yovanovich Eq. (\ref{eq:rspr}))\cite{yovanovich2002analytical,muzychka1999thermal}
    \item $R_{\text{BD}} = 1/(h_{\text{BD}} \cdot \pi b^2)$: Interface boundary resistance
    \item $f = [b/(b + L_{\text{spread}})]^2$: Geometric dilution factor
    \item $L_{\text{spread}}$: Lateral spreading length
\end{itemize}

The impact of the interfacial thermal resistance ($R_{int}$) and its physically reasonable range are determined using FEM modeling, as discussed in Sec.~\ref{sec:results}. Furthermore, heat-spreading analysis is crucial for determining the thermal conductivity ($k$) of the sample. Yovanovich \emph{et al.}\cite{yovanovich2002analytical,yovanovich2005four} and Muzychka \emph{et al.}\cite{muzychka1999thermal} derived analytical solutions for the thermal spreading resistance ($R_{spr}$), as shown in Eq.~(\ref{eq:rspr}), for a film on a substrate as a function of the film thickness ($t$) and the thermal conductivities of the film ($k_{eff}$) and substrate ($k_{sub}$).
\begin{equation}
R_{spr} = \frac{1}{\pi k_{eff} b} 
\int_{0}^{\infty} 
\left[ \frac{1 + K \exp(-2\xi t / b)}{1 - K \exp(-2\xi t / b)} \right] 
J_1(\xi) \, \sin(\xi) \, \frac{d\xi}{\xi^2}
\label{eq:rspr}
\end{equation}
where $K$ is defined as $K=\frac{1-k_{sub}/k_{eff}}{1+k_{sub}/k_{eff}}$, $b$ is the source radius, and $J_1$ is the first-order Bessel function of the first kind. 

The buried interface thermal boundary conductance $h_{\text{BD}}$ is a critical parameter in this analysis. However, as heat transfers through a thin film to the substrate, it spreads laterally before reaching the buried interface. The characteristic lateral spreading length ($L_{\text{spread}}$) determines the effective area over which heat crosses the buried interface. For heat flow from a localized source of radius $b$ at the top of a thin film of thickness $t$ and thermal conductivity $k_{\text{film}}$, deposited on a substrate with thermal conductivity $k_{\text{sub}}$, this spreading length is given by Eq.~(\ref{eq:Lspread})\cite{muzychka2004thermal,yovanovich2003thermal,kennedy1960spreading}.
\begin{equation}
    L_{\text{spread}} = t \cdot \sqrt{\frac{k_{\text{film}}}{k_{\text{sub}}}}
    \label{eq:Lspread}
\end{equation}
This lateral spreading reduces the effective contribution of the buried interface resistance through a geometric dilution effect. The dilution factor $f$ represents the fraction of interface resistance ``seen'' by the heat flux and is given by Eq.~(\ref{eq:dilution}).
\begin{equation}
f = \left(\frac{b}{b + L_{\text{spread}}}\right)^2 = \left(\frac{b}{b + t\sqrt{k_{\text{film}}/k_{\text{sub}}}}\right)^2
\label{eq:dilution}
\end{equation}
Physically, heat spreading reduces the effective contribution of interface resistance by the ratio of the original contact area to the expanded area at the interface. As the spreading length increases relative to the contact radius, more heat bypasses the localized interface resistance, thereby reducing its impact on the total thermal resistance.

A critical insight is that both $R_{\text{spr}}(k)$ and $f$ depend on $k_{\text{film}}$, which is the unknown quantity we seek to determine. This creates a self-consistency requirement: the spreading resistance $R_{\text{spr}}(k)$ and the effective buried interface resistance $f(k) \cdot R_{\text{BD}}$ are both functions of the unknown film thermal conductivity. Consequently, the thermal resistance balance (Eq.~\ref{eq:selfconsistent})
\begin{equation}
R_{\text{spr}}(k) + f(k) \cdot R_{\text{BD}} = (R_c - R_{nc})_{meas} - R_{\text{int}}
\label{eq:selfconsistent}
\end{equation}
must be solved iteratively to extract $k_{\text{film}}$. This iterative solution converges to yield the thermal conductivity of the thin film, with the complete algorithm detailed in the section S2. (Supplementary Material).

\begin{figure}
\includegraphics{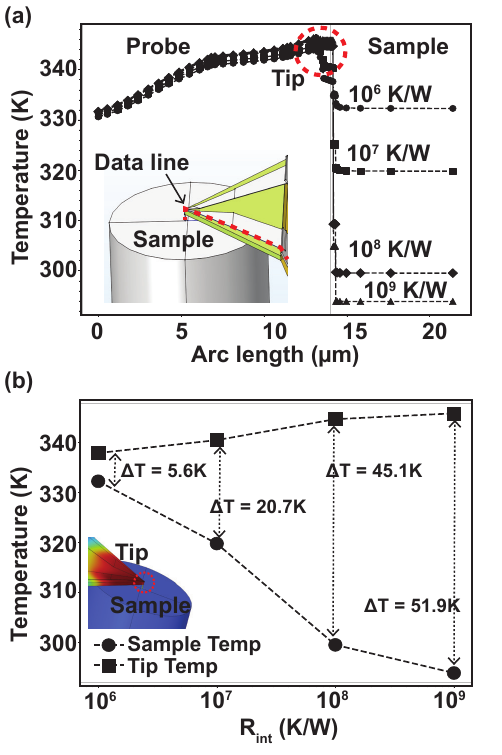}
\caption{\label{fig:rint_temp_profile} (a) Temperature distribution along the arc length of data line for different thermal interface resistances ($R_{int}$). Data points are downsampled for clarity, with distinct markers and dotted lines highlighting the temperature trends for each $R_{int}$ value. (b) Temperature difference ($\Delta T = T_{tip} - T_{sample}$) versus $R_{int}$, illustrating the monotonic increase of thermal decoupling with higher contact resistance. Together, these plots highlight the influence of $R_{int}$ on probe–sample thermal coupling in SThM measurements.}
\end{figure}

\section{\label{sec:results}Results and Discussion}
\subsection{Finite-element multiphysics modeling of SThM probe}  
The variation in the working environment of SThM measurements and the uncertainty in the thermal resistance of the tip-sample interface ($R_{int}$) make quantitative measurements challenging. Therefore, a fundamental understanding of thermal transport at the tip-sample interface is essential for the quantitative determination of local thermal conductivities of the sample.  In this study, a detailed three-dimensional finite element model (3D-FEM) is developed to evaluate the interfacial thermal resistance ($R_{int}$) by simulating the heat transfer mechanisms occurring at the probe–sample junction. Accurate determination of $R_{int}$ provides insights into the complex heat transfer pathways occurring between the tip and the sample. This 3D-FEM model is developed using COMSOL Multiphysics\textsuperscript{\textregistered} to systematically explore the effects of multiple physical parameters on the tip–sample thermal interaction. Fig.~\ref{fig:rint_surface}(a) illustrates the SThM probe and sample structure, along with their meshed geometry. The probe comprises a Si$_3$N$_4$ cantilever, an Au pad connected to the tip, and a Pd tip. Detailed dimensions and structural specifications of probe are provided in Fig. S5 (Supplementary Material). The model incorporates two primary physical processes: (1) heat transfer in solids and (2) electromagnetic heating (Joule heating in the Pd resistor). All exposed surfaces are set as adiabatic, except for the interface between the tip and the sample.

The quasi-equilibrium heat transfer equation with internal Joule heating is employed to model heat transport throughout the system:
\begin{equation}
- \nabla \cdot \left( \kappa \nabla T \right) = Q_e
\label{eq:fourier's law}
\end{equation}
where $\kappa$ is the thermal conductivity. The volumetric Joule heat source $Q_e$ arises from electrical biasing of the Pd thermistor and is given by
\begin{equation}
Q_e = \frac{V_{p}^2}{\Omega(T) \mathcal{V}_h}
\label{eq:joule}
\end{equation}
Where $V_p$ is the applied voltage, $\Omega(T)$ is the temperature-dependent electrical resistance of the Pd thermistor, and $\mathcal{V}_h$ is the effective heater volume. This expression corresponds to uniform distribution of electrical power dissipation within the resistive element. Convection term was not considered in Eqs.~(\ref{eq:fourier's law}) due to its negligible contribution under the given experimental conditions \cite{zhang2023realizing, bodzenta2022scanning, deshmukh2022direct}. 

The heat generated at the probe tip is transferred through the tip–sample interface and subsequently into the sample. To accurately capture the variation in $R_{int}$, the SThM tip–sample contact is modeled over a wide range of $R_{int}$ values, from $10^6$ to $10^9$ K/W, as illustrated in Fig.~\ref{fig:rint_temp_profile}(a). This simulation range is selected based on experimentally obtained thermal resistance values for nanoscale contacts\cite{li2025quantitative,gomes2015scanning}. At low interfacial resistance ($R_{int} < 10^6$ K/W), the contribution of $R_{int}$ to the overall thermal resistance ($R_c$) is negligible, indicating perfect heat transfer across the interface. On the other hand, at high interfacial resistance ($R_{int} > 10^9$ K/W), heat transfer between the tip and sample becomes significantly suppressed, effectively isolating the thermal transport pathways. The resulting probe temperature profiles for different $R_{int}$ values are presented in Fig.~\ref{fig:rint_temp_profile}(a). At high interface thermal resistance values ($R_{int} \geq 10^8\text{K/W}$), the sample temperature remains constant at room temperature, indicating the absence of local heating. Therefore, $R_{int}$ cannot exceed $10^7 \text{K/W}$, since local heating of the sample is need for thermal resistance measurement.

The effect of the $R_{int}$ on local sample heating was examined for two materials with strongly contrasting intrinsic thermal conductivities: Cu (300 W/m$\cdot$K) and SiO$_2$ (1.38 W/m$\cdot$K). As shown in Fig.~\ref{fig:rint_surface}(b–d) for SiO$_2$ and Fig.~\ref{fig:rint_surface}(f–h) for Cu, identical values of 
$R_{int} $ result in qualitatively similar surface temperature distributions beneath the tip, despite the large difference in bulk thermal conductivity. Because the COMSOL model considers smooth surfaces for all domains, the resulting temperature profiles for Cu and SiO$_2$ are expected to be identical, which is in agreement with previous studies showing that $R_{int}$ is independent of the underlying material in the absence of surface roughness effects \cite{li2025quantitative,pernot2021frequency, bahrami2004thermal}.  At lower interfacial thermal resistance ($R_{int} \leq 10^6$ K/W), efficient heat transfer from the probe leads to a broader thermal diffusion profile in both materials. In contrast, increasing 
$R_{int} \geq 10^7$ K/W strongly suppresses heat flow across the interface, resulting in negligible local surface heating for both Cu and SiO$_2$. Importantly, the experimentally measured total thermal resistance lies in the range $ \leq 10^7$ K/W and is accompanied by clear local surface heating beneath the probe. Therefore, 
 $R_{int}$ must be 
$\leq 10^7$ K/W and cannot exceed the experimentally inferred thermal resistance. 

Furthermore, the difference between the SThM tip temperature and the corresponding local sample temperature variation is plotted in Fig.~\ref{fig:rint_temp_profile}(b). The change in tip and sample temperature ($\Delta T$) is shown for various $R_{int}$ values.  The results shown in Fig.~\ref{fig:rint_temp_profile}(a,b) demonstrate that $R_{int}$ plays a dominant role in the quantitative determination of the sample’s local thermal conductivity. While $R_{int} $ is treated parametrically in the FEM simulations to examine its influence on the measured temperature response, this approach is adopted to establish a physically constrained relationship between the experimentally accessible temperature contrast $\Delta T=T_{tip} - T_{sample}$ and $R_{int}$. By comparing the experimentally measured $\Delta T$ (Fig.~\ref{fig:Rp}) with the trends predicted by the FEM model shown in Fig.~\ref{fig:rint_temp_profile} (b), a plausible and bounded range of $R_{int} $ is identified for the given probe–sample contact conditions.

\begin{figure*}[htbp]
\includegraphics{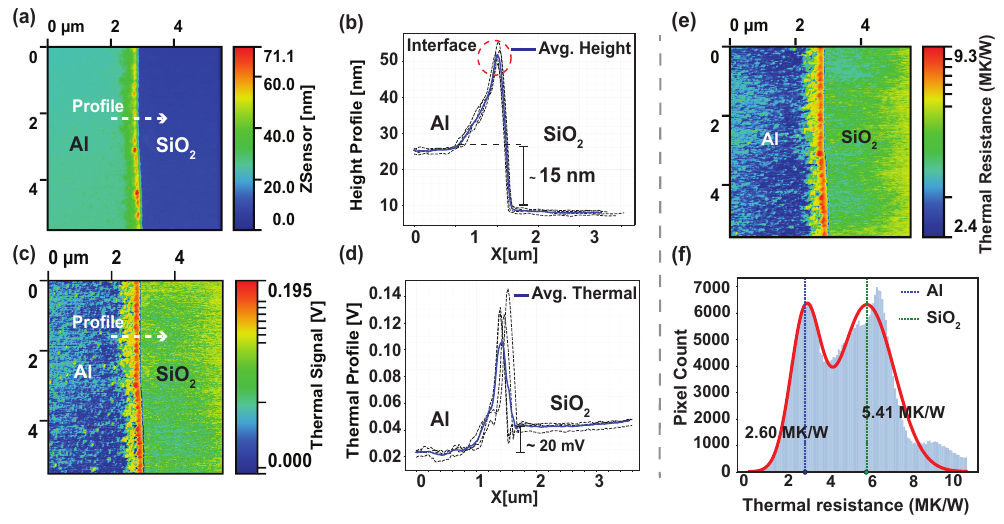}% Here is how to import EPS art
\caption{\label{fig:scan_surface} (a) AFM topography image of a $5~\mu\mathrm{m} \times 5~\mu\mathrm{m}$ region of the Al/SiO$_2$ sample surface. (b) Corresponding height profile across the interface, showing a step height of 15\,nm for the Al  layer, consistent with the sample fabrication procedure. (c) SThM thermal signal map, where the lower heating voltage over Al  indicates higher thermal conductivity compared to SiO$_2$. (d) Line profile of the thermal signal across the Al/SiO$_2$ interface, revealing elevated heating voltage at the boundary due to topography artifacts. A relative thermal signal difference of ~20 mV is observed between Al  and SiO$_2$ surfaces.(e) Thermal signal map converted to thermal resistance ($R_c$) map through calibrated mathematical modeling. (f) 1D histogram derived from pixel-to-pixel correlation of thermal signals, revealing prominent thermal resistances.
}
\end{figure*}
\subsection{Experimental determination of model parameters}

To experimentally validate the quantitative SThM model, a 15\,nm aluminum (Al) thin film deposited on a SiO$_2$ substrate was selected as the test sample. The film was patterned adjacent to the bare SiO$_2$ substrate, enabling direct comparison between regions of differing thermal conductivity under identical measurement conditions. The sample preparation and fabrication details are provided in the Sec. \ref{sec:methodology} (A). All SThM measurements were carried out at a controlled room temperature of about 294 K, monitored continuously by a temperature regulation system. The SThM probe was positioned near the Al/SiO$_2$ interface and brought into mechanical contact with the sample surface. This contact led to a sudden reduction in probe temperature, indicating the activation of an additional thermal conduction pathway at the tip–sample junction, thereby enhancing heat dissipation into the sample (see Fig. S1 from the Supplementary Material). Subsequently, as the tip scanned the sample area, the probe recorded thermal signals from the sample as temperature variations, which allowed the calculation of the sample's thermal resistance. A total of 7 independent SThM scans were acquired to ensure the reproducibility of the measurements. The corresponding pixel-level thermal signal and topography data were extracted using the SPM image analysis software named Gwyddion. The data were processed and analyzed using a custom-developed Python script designed specifically for quantitative pixel-wise thermal signal correlation and noise reduction. This in-house code enabled automated data extraction, normalization, and statistical comparison across multiple scans, significantly improving the accuracy and consistency of the thermal mapping results.

Figure. \ref{fig:scan_surface}(a,c) shows the topography and thermal signal map of 15\,nm-thick Al thin film patterned on a SiO$_2$ substrate. A strong thermal signal contrast was observed between the Al film and the SiO$_2$ substrate, reflecting the significant differences in their thermal conductivities. Two approach-retraction curves for either side of the interface  are used for the conversion between the SThM signal and thermal resistance using Eq. (\ref{eq:voltagethermalresistance}). Further details about the approach-retraction curve are provided in Fig. S1 (Supplementary Material). The relative change in the thermal signal ($\sim$20 mV) from the SiO$_2$ surface to the Al film corresponds to a decrease in thermal resistance from the SiO$_2$ surface to the 15\,nm-thick Al film. Therefore, this decrease in thermal resistance indicates a higher thermal conductivity of the Al film compared to the SiO$_2$ substrate. 

To extract the thermal resistance associated with Al and SiO$_2$, a pixel-to-pixel correlation between the height and thermal signal maps is performed. This correlation is used to construct a 2D histogram (see Fig. S2 from the Supplementary Material), which enables a separation of the thermal responses from Al and SiO$_2$ surfaces. Traditional analysis methods often rely on arbitrarily selected line-cuts or cropped regions of interest, introducing potential user bias in the selection process. In contrast, the present approach systematically processes all available data points from the thermal signals measurement, ensuring reproducibility and minimizing subjective interpretation. This comprehensive analysis leverages the full spatial information content, thereby improving statistical confidence in the extracted thermal properties. Subsequently, the acquired data are subjected to statistical processing to extract thermal properties. However, this analysis is susceptible to errors arising from experimental artifacts—such as surface contamination, tip-sample contact variations, or instrumental noise—which can distort the thermal signal and compromise measurement accuracy. Therefore, to quantitatively extract representative thermal resistance values while accounting for spatial heterogeneity, a 2D histogram of thermal resistance on the y-axis and film thickness on the x-axis is constructed (Fig.~S2 of Supplementary Material). This histogram is then projected onto the thermal resistance axis to obtain statistically representative values. This projection combines measurements across all thickness values, yielding a one-dimensional distribution from which statistically robust resistance values can be determined. The resulting 1D distribution is modeled using a double-Gaussian function, from which the mean peak positions corresponding to the thermal resistances of Al and SiO$_2$ are extracted. The statistical uncertainties in these mean values are obtained from the covariance matrix of the nonlinear least-squares fitting procedure. From this analysis, the mean thermal resistances were found to be 2.60 $\pm$ 0.03 MK/W and 5.41 $\pm$ 0.04 MK/W  for Al and SiO$_2$, respectively. 

The measured thermal resistances $R_{\text{meas}}(\text{SiO}_2) = 5.41 \pm 0.04$~MK/W and $R_{\text{meas}}(\text{Al}) = 2.60 \pm 0.03$~MK/W provide the experimental foundation for quantitative determination of the thermal conductivity of the Al thin film. Following the theoretical framework established in Sec.~\ref{sec:methodology}, we model the measured contact-to-noncontact thermal resistance difference as a series combination of three contributions: the tip-sample interfacial thermal resistance ($R_{\text{int}}$), the sample spreading resistance ($R_{\text{spr}}$), and the Al/SiO$_2$ buried interface thermal boundary resistance ($R_{\text{BD}}$), as given by Eq.~(\ref{eq:seriesrth}). This formulation enables direct extraction of the Al thermal conductivity from the experimental data through an iterative algorithm (Eq.~\ref{eq:selfconsistent}) that accounts for the interdependence of $R_{\text{spr}}$ and $R_{\text{BD}}$ on the unknown film thermal conductivity.
\subsubsection{Tip Interface Resistance  $R_{int}$ Extraction}
As shown in Fig.~\ref{fig:scan_surface}(e), the experimentally measured $(R_c - R_{nc})_{meas}$ values are $\leq 10^7$ K/W. Therefore, it can be concluded that $R_{int}$ is also $\leq 10^7$ K/W. In addition, the tip–sample contact resistance ($R_{int}$) is mainly governed by surface roughness, the applied tip–sample normal force, and the resulting contact area \cite{pernot2021frequency, bahrami2004thermal,puyoo2010thermal}. Therefore, the thermal contact between the SThM probe and the sample may slightly differ for Al and SiO$_2$ surfaces; however, it is very challenging to determine the exact $R_{int}$ for each case. Therefore, for simplification, we assumed the same $R_{int}$ for both Al and SiO$_2$ interfaces. This assumption introduces negligible uncertainty in the measurement because both surfaces exhibit comparable roughness (see Fig. S6 in the Supplementary Material). Additionally, the tip-sample normal contact force was maintained with an AFM setpoint value of $\sim$0.6 V, which corresponds to < 10 nN contact force for this type of probes. Therefore, the tip-sample interface resistance $R_{\text{int}}$ is extracted from a reference measurement on bare SiO$_2$/Si (no thin film, $R_{\text{BD}} \approx 0$).

\begin{equation}
    R_{\text{int}} = R_{\text{meas}}(\text{SiO}_2) - R_{\text{spr}}(\text{SiO}_2/\text{Si})
    \label{eq:Rint_extraction}
\end{equation}

where $R_{\text{spr}}(\text{SiO}_2/\text{Si})$ is calculated using Eq.~(\ref{eq:rspr}) with known $k_{\text{SiO}_2}$ (1.38 W/m$\cdot$K) and $k_{\text{Si}}$ (150 W/m$\cdot$K). The parameters used in this analysis are provided in Table S1 (supplementary material).
\subsubsection{Iterative Algorithm for $k$ Extraction}
 Literature values for metal/dielectric interfaces\cite{stoner1993kapitza,lombard2014influence,schmidt2008pulse,cheaito2015thermal} span a considerable range. For the Al/SiO$_2$ interface, we adopt a central value of $h_{\text{BD}} = 200$~MW/(m$^2$K) with a realistic range of $h_{\text{BD}} = 150$--250~MW/(m$^2$K) to account for this uncertainty \cite{hopkins2010criteria, kwon2021thermal,zhu2010ultrafast}.

Applying the iterative algorithm to the experimental measurements yields a thermal conductivity of $k_{\text{Al}} = 45.1$~W/(m$\cdot$K) at the central values of $h_{\text{BD}} = 200$~MW/(m$^2$K) and $t = 14$~nm. See Section S3 (supplementary material) for the detailed calculation. In our 15~nm Al film, a sub-stoichiometric Al$_2$O$_{3-x}$ layer ($\sim$1--2~nm thick) forms naturally on the Al surface due to exposure to ambient air\cite{yang2018liquid,gorobez2021growth}, reducing the effective conductive metallic thickness to approximately 13--14~nm.  A comprehensive uncertainty analysis accounting for variations in both interface conductance ($h_{\text{BD}} = 150$--250~MW/(m$^2$K)) and film thickness ($t = 13$--15~nm) produces asymmetric error bounds of $+4.7/-3.6$~W/(m$\cdot$K), yielding a final result of $k_{\text{Al}} = 45.1^{+4.7}_{-3.6}$~W/(m$\cdot$K).  This represents a 5.3-fold reduction from bulk aluminum (237~W/(m$\cdot$K)), consistent with Fuchs-Sondheimer theory\cite{fuchs1938conductivity,Sondheimer1952MeanFreePath} for films of thickness comparable to the electron mean free path ($t/\lambda_e \approx 0.93$). The thermal resistance budget (Table~\ref{tab:thermal_resistance}) reveals that the tip-sample interface dominates the total measured resistance (70.8\%), with spreading resistance contributing 26.2\% and the diluted buried interface resistance only 3.1\%. The geometric dilution factor of $f = 0.148$ indicates that lateral heat spreading over a characteristic length of $L_{\text{spread}} \approx 80$~nm substantially reduces the effective contribution of the Al/SiO$_2$ interface resistance. A complete matrix of extracted thermal conductivity values for all combinations of $h_{\text{BD}}$ and thickness is provided in the Table. S1 of Supplementary Material.
\begin{table}
    \caption{\textbf{Thermal resistance budget breakdown.}}
    \label{tab:thermal_resistance}
    \begin{ruledtabular}
    \begin{tabular}{lccc}
        Component & Value (MK/W) & Percentage & Description \\
        \hline
        $R_{\text{int}}$ (tip-Al) & 1.84 & 70.8\% & Tip-sample interface \\
        $R_{\text{spr}}$ (Al/SiO$_2$) & 0.68 & 26.2\% & Spreading resistance \\
        $f \cdot R_{\text{BD}}$ (Al/SiO$_2$) & 0.08 & 3.1\% & Diluted  $R_{\text{BD}}$ \\
        \hline
       \textbf{Total} & \textbf{2.60} & \textbf{100}\% & \textbf{=$R_{\text{meas}}$} \\
    \end{tabular}
    \end{ruledtabular}
\end{table}

\begin{table}[htbp]
\caption{Phonon thermal conductivity ($k_p$), electron thermal conductivity ($k_e$), and total thermal conductivity ($k$) of bulk aluminum from literature}
\label{tab:thermal_conductivity}
\begin{ruledtabular}
\begin{tabular}{l c c c}
Metal &
$k_p$ (W/m$\cdot$K) &
$k_e$ (W/m$\cdot$K) &
$k$ (W/m$\cdot$K) \\
\hline
Al &
6 &
246 &
237--252 \\
\end{tabular}
\end{ruledtabular}
\footnotesize Data adapted from Refs.~\cite{jain2016thermal, kittel2018introduction}.
\end{table}

\begin{figure}[htbp]
\includegraphics[width=0.5\textwidth]{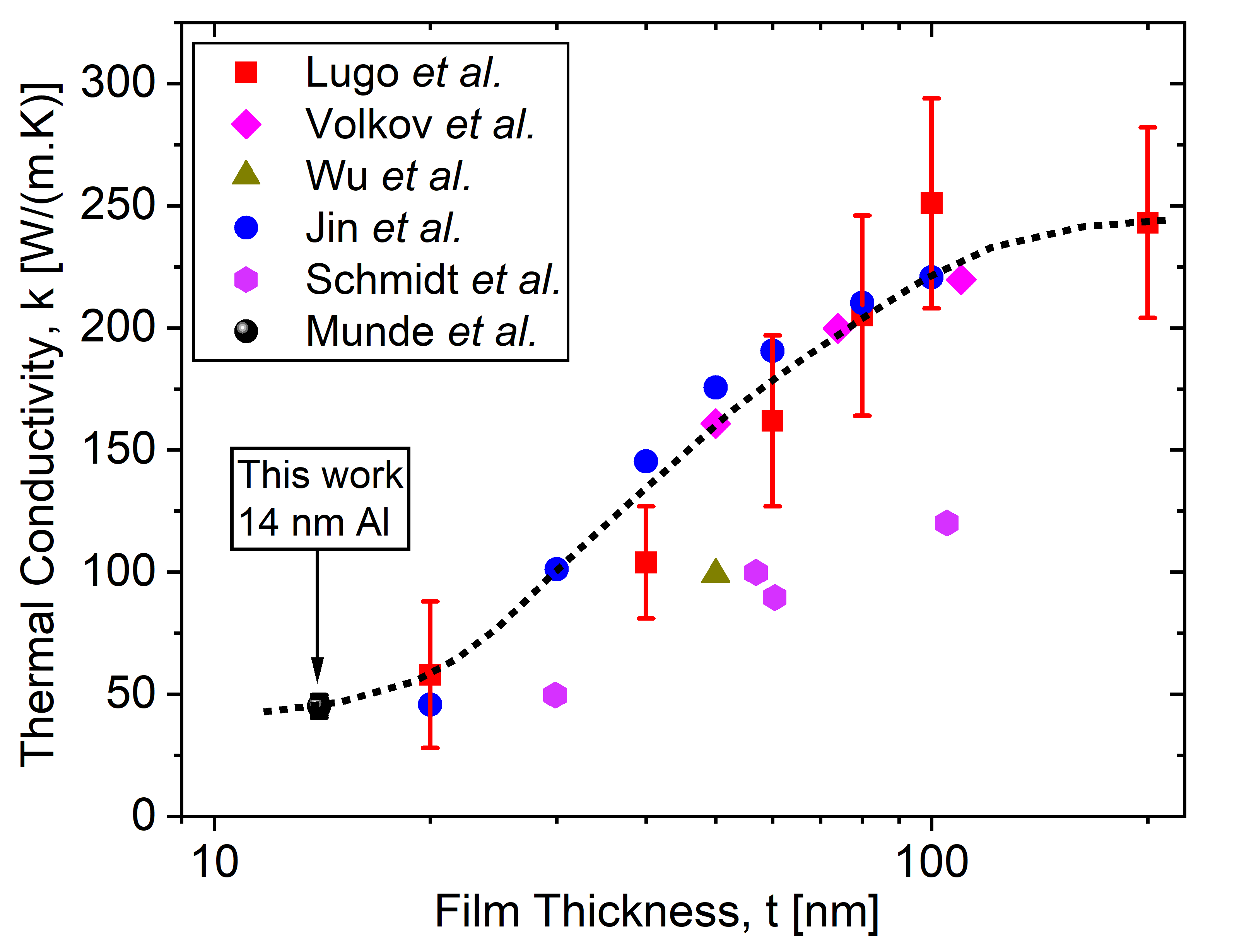}% Here is how to import EPS art
 \caption{Thickness-dependent thermal conductivity of polycrystalline Al thin films. The SThM result from this work (14~nm Al, $k =45.1^{+4.7}_{-3.6}$~W/(m$\cdot$K)) is benchmarked against reported values from Lugo \textit{et al.}\cite{lugo2016thermal}, Volkov \textit{et al.}\cite{Volkov1976AluminumThermal}, 
Wu \textit{et al.}\cite{Wu1993PhotothermalThinFilms}, Jin \textit{et al.}\cite{Jin2008ElectronMeanFreePath}, and Schmidt \textit{et al.}\cite{Schmidt2010ThinMetalFDTR} The dashed curve is a trend line, consistent with the Fuchs-Sondheimer\cite{fuchs1938conductivity, Sondheimer1952MeanFreePath} model incorporating Mayadas-Shatzkes\cite{mayadas1970electrical} grain boundary scattering. Data spread reflects variations in film microstructure and measurement techniques.}
    \label{fig:literature}
\end{figure}
\section*{Discussion}

The thermal conductivity of Al consists of both phonon ($k_p$) and electron ($k_e$) contributions, with electrons typically dominating heat transport in bulk. Jain \textit{et al.} performed detailed first-principles calculations to examine both phonon-phonon and electron-phonon scattering processes in bulk Al, quantitatively determining the individual contributions of $k_p$ and $k_e$ to the overall thermal conductivity\cite{jain2016thermal}. Furthermore, they quantified the mean free path ($\Lambda$) distributions of both phonon and electron heat carriers by analyzing their respective thermal conductivity accumulation functions at 300~K (see Fig. S7 of Supplementary Material). Analysis of the thermal conductivity accumulation function demonstrates that the electron contribution decreases substantially as $\Lambda$ decreases, indicating that electrons with short $\Lambda$ contribute less to heat transport than those with longer $\Lambda$. 

The accumulation function data at 300~K indicate that phonon thermal conductivity in Al is dominated by carriers with mean free paths ranging from 1~nm to 22~nm (comprising $\sim$90\% of $k_p$), while electron thermal conductivity requires longer mean free paths of 10~nm to 25~nm for similar accumulation ($\sim$90\% of $k_e$). The onset of contributions also differs markedly: phonon transport begins at $\Lambda \sim 1$~nm, whereas electron transport only becomes substantial at $\Lambda > 10$~nm. These characteristic length scales directly impact thermal transport in confined geometries. This confinement preferentially suppresses electron transport, as the film dimension is smaller than the typical electron mean free paths, resulting in diffuse boundary scattering that significantly reduces $k_e$. 

%The extracted thermal conductivity of $k_{\text{Al}} = 45.1^{+4.7}_{-3.6}$~W/(m$\cdot$K) represents a 5.3-fold reduction from bulk aluminum (237~W/(m$\cdot$K)). This result satisfies all physical consistency checks. First, it exceeds the phonon-only thermal conductivity floor of aluminum ($\sim$6~W/(m$\cdot$K))\cite{jain2016thermal,kittel2018introduction}, confirming that electron transport remains active despite strong boundary scattering. Second, our value falls within the range predicted by Fuchs-Sondheimer theory\cite{fuchs1938conductivity} for films with $t/\lambda_e \approx 0.93$, which yields $k/k_{\text{bulk}} \approx 0.15$--0.25, corresponding to $k \approx 35$--60~W/(m$\cdot$K). Third, published data for Al thin films at similar thicknesses ($t \approx 20$~nm) report $k \approx 50$--60~W/(m$\cdot$K)\cite{lugo2016thermal,zhang2006influence,nath1974thermal}, consistent with our slightly thinner film exhibiting somewhat lower conductivity. Fourth, the Mayadas-Shatzkes model\cite{mayadas1970electrical} predicts additional reduction due to electron scattering at grain boundaries, which explains why our value lies at the lower end of the Fuchs-Sondheimer prediction range. Fig.7(a,b) compares our result with published data for metal thin films, showing excellent agreement on thickness-dependent and a universal scaling plot of $k/k_{\text{bulk}}$ versus $t/\lambda_e$.
Fig.~\ref{fig:literature} compares the SThM-extracted thermal conductivity of the 14~nm Al film ($k = 45.1^{+4.7}_{-3.6}$~W/(m$\cdot$K)) with literature values for polycrystalline Al thin films of varying thickness measured by different techniques, including frequency-domain thermoreflectance (FDTR)~\cite{Schmidt2010ThinMetalFDTR}, photothermal mirage~\cite{Wu1993PhotothermalThinFilms}, electrical heating with temperature profile analysis~\cite{lugo2016thermal}, and electron diffraction thermometry~\cite{Volkov1976AluminumThermal}. Boltzmann transport equation (BTE) modeling results from Jin \textit{et al.}~\cite{Jin2008ElectronMeanFreePath} are also included. Our result falls on the expected trend of decreasing thermal conductivity with decreasing film thickness, consistent with enhanced electron scattering at surfaces and grain boundaries.

The dashed curve in Fig.~\ref{fig:literature} is a trend line that captures the thickness-dependent behavior and remains consistent with the combined Fuchs-Sondheimer and Mayadas-Shatzkes electron scattering model~\cite{fuchs1938conductivity,Sondheimer1952MeanFreePath,mayadas1970electrical}. Using the literature value of $\lambda_e = 19$~nm for the electron mean free path in aluminum~\cite{Gall2016}, this theoretical framework predicts $k = 49.5$~W/(m$\cdot$K) for a 14~nm film (Supplementary Section~S4), which falls within our experimental uncertainty range. The spread in literature data reflects variations in film microstructure, deposition conditions, and measurement techniques across different studies. The agreement between our SThM measurement, the theoretical prediction, and existing literature data validates the iterative extraction methodology developed in this work.

The thermal signal and topography appear elevated along the Al/SiO$_2$ boundary (see Fig. 6(a, c)). The elevated topography introduces a minor artifact in the thermal signal due to AFM feedback sensitivity effects at the step height. Consequently, the higher thermal resistance extracted at the edge arises from the same artifact, as it is directly derived from the thermal signal. Nevertheless, Fig. 6(e) demonstrates the significant potential of this approach for probing thermal transport variations across dislocation defects and grain boundaries with sub-100~nm spatial resolution. Furthermore, validation using established approaches, such as those described in Refs.~\onlinecite{tiwari2013anisotropic, isotta2023microscale} for assessing thermal property variations across grain boundaries with different densities, would be highly beneficial, particularly given the high spatial resolution offered by SThM.

\section{Conclusion}
We experimentally quantified the thermal resistance of the SThM probe using a circuit-based calibration method and employed FEM modeling to investigate the underlying heat transfer mechanisms. The contributions of various probe–sample heat transfer channels, along with their corresponding measurement strategies, were systematically examined. Furthermore, the thermal conductivity of a 15\,nm Al thin film, deposited by e-beam evaporation on a few-micrometer-thick SiO$_2$ substrate, is measured. The robustness of the pixel-to-pixel correlation approach for extracting representative thermal signals is discussed. In addition, statistical and measurement uncertainties in thermal resistance are quantified, considering the underlying assumptions. The measured thermal conductivity of the Al thin film is in good agreement with the theoretical frameworks and the published experimental data.

Overall, as demonstrated in this study, the SThM-based thermal metrology approach enables quantitative and spatially resolved characterization of thermal transport at the nanoscale. This technique provides sub-100\,nm lateral resolution, allowing direct mapping of local heat conduction pathways with precision far beyond that achievable by conventional optical methods such as TDTR and FDTR. Unlike TDTR and FDTR, which are fundamentally constrained by optical diffraction limits and thermal penetration depth—thereby limiting their reliability for ultrathin films—SThM imposes no such thickness restrictions. This advantage makes it particularly effective for studying two-dimensional materials, nanoscale heterostructures, and complex interfaces. SThM emerges as a powerful and complementary tool for nanoscale thermal transport characterization, offering insights into defect-, boundary-, and interface-driven effects that are inaccessible through traditional macroscopic or optical approaches.

\section*{Supplementary Material}
See the supplementary material for the raw data that supports the main findings of this study.

\begin{acknowledgments}
 We acknowledge Dr. Amy Marconnet, Professor of Mechanical Engineering at Purdue University, for her valuable guidance and for providing access to the COMSOL license.
\end{acknowledgments}

\section*{Data Availability Statement}

The experimental and modeling data that support the findings of this study are available from the corresponding author upon reasonable request.

\section*{References}
%\nocite{*}

\bibliography{aipsamp}% Produces the bibliography via BibTeX

@article{mcconnell2005thermal,
  title={Thermal conduction in silicon micro-and nanostructures},
  author={McConnell, AD and Goodson, Kenneth E},
  journal={Annual review of heat transfer},
  volume={14},
  year={2005},
  publisher={Begel House Inc.}
}

@article{cahill1987thermal,
  title={Thermal conductivity of amorphous solids above the plateau},
  author={Cahill, David G and Pohl, Robert O},
  journal={Physical review B},
  volume={35},
  number={8},
  pages={4067},
  year={1987},
  publisher={APS}
}

@article{cahill2002thermometry,
  title={Thermometry and thermal transport in micro/nanoscale solid-state devices and structures},
  author={Cahill, David G and Goodson, Kenneth and Majumdar, Arunava},
  journal={J. Heat Transfer},
  volume={124},
  number={2},
  pages={223--241},
  year={2002}
}

@article{majumdar1993thermal,
  title={Thermal imaging using the atomic force microscope},
  author={Majumdar, A and Carrejo, JP and Lai, J},
  journal={Applied Physics Letters},
  volume={62},
  number={20},
  pages={2501--2503},
  year={1993},
  publisher={American Institute of Physics}
}

@article{bodzenta2022scanning,
  title={Scanning thermal microscopy and its applications for quantitative thermal measurements},
  author={Bodzenta, Jerzy and Ka{\'z}mierczak-Ba{\l}ata, Anna},
  journal={Journal of Applied Physics},
  volume={132},
  number={14},
  year={2022},
  publisher={AIP Publishing}
}

@article{li2022comprehensive,
  title={A comprehensive review for micro/nanoscale thermal mapping technology based on scanning thermal microscopy},
  author={Li, Yifan and Zhang, Yuan and Liu, Yicheng and Xie, Huaqing and Yu, Wei},
  journal={Journal of Thermal Science},
  volume={31},
  number={4},
  pages={976--1007},
  year={2022},
  publisher={Springer}
}

@article{gucmann2021scanning,
  title={Scanning thermal microscopy for accurate nanoscale device thermography},
  author={Gucmann, Filip and Pomeroy, James W and Kuball, Martin},
  journal={Nano Today},
  volume={39},
  pages={101206},
  year={2021},
  publisher={Elsevier}
}

@article{li2025quantitative,
  title={Quantitative measurements in scanning thermal microscopy: Theoretical models, calibration technique, and integrated instrument},
  author={Li, Yifan and Wu, Jiang and Luo, Jing and Wang, Jianli and Yu, Wei and Cao, Bingyang},
  journal={Journal of Applied Physics},
  volume={138},
  number={5},
  year={2025},
  publisher={AIP Publishing}
}

@article{huang2024violation,
  title={Violation of the Wiedemann--Franz Law and Ultralow Thermal Conductivity of Ti3C2T x MXene},
  author={Huang, Yubin and Spiece, Jean and Parker, Tetiana and Lee, Asaph and Gogotsi, Yury and Gehring, Pascal},
  journal={ACS nano},
  volume={18},
  number={47},
  pages={32491--32497},
  year={2024},
  publisher={ACS Publications}
}

@article{razeghi2023single,
  title={Single-material MoS2 thermoelectric junction enabled by substrate engineering},
  author={Razeghi, Mohammadali and Spiece, Jean and O{\u{g}}uz, O{\u{g}}uzhan and Pehlivano{\u{g}}lu, Doruk and Huang, Yubin and Sheraz, Ali and Ba{\c{s}}{\c{c}}{\i}, U{\u{g}}ur and Dobson, Phillip S and Weaver, Jonathan MR and Gehring, Pascal and others},
  journal={npj 2D Materials and Applications},
  volume={7},
  number={1},
  pages={36},
  year={2023},
  publisher={Nature Publishing Group UK London}
}

@article{gomes1999dc,
  title={DC thermal microscopy: study of the thermal exchange between a probe and a sample},
  author={Gom{\`e}s, S{\'e}verine and Trannoy, Nathalie and Grossel, Philippe},
  journal={Measurement Science and Technology},
  volume={10},
  number={9},
  pages={805},
  year={1999},
  publisher={IOP Publishing}
}

@article{luo1997sensor,
  title={Sensor nanofabrication, performance, and conduction mechanisms in scanning thermal microscopy},
  author={Luo, K and Shi, Z and Varesi, J and Majumdar, A},
  journal={Journal of Vacuum Science \& Technology B: Microelectronics and Nanometer Structures Processing, Measurement, and Phenomena},
  volume={15},
  number={2},
  pages={349--360},
  year={1997},
  publisher={American Vacuum Society}
}

@article{zhang2020review,
  title={A review on principles and applications of scanning thermal microscopy (SThM)},
  author={Zhang, Yun and Zhu, Wenkai and Hui, Fei and Lanza, Mario and Borca-Tasciuc, Theodorian and Mu{\~n}oz Rojo, Miguel},
  journal={Advanced functional materials},
  volume={30},
  number={18},
  pages={1900892},
  year={2020},
  publisher={Wiley Online Library}
}

@article{assy2015temperature,
  title={Temperature-dependent capillary forces at nano-contacts for estimating the heat conduction through a water meniscus},
  author={Assy, Ali and Gom{\`e}s, S{\'e}verine},
  journal={Nanotechnology},
  volume={26},
  number={35},
  pages={355401},
  year={2015},
  publisher={IOP Publishing}
}

@article{yovanovich2002analytical,
  title={Analytical modeling of spreading resistance in flux tubes, half spaces, and compound disks},
  author={Yovanovich, M Michael and Culham, J Richard and Teertstra, Pete},
  journal={IEEE Transactions on Components, Packaging, and Manufacturing Technology: Part A},
  volume={21},
  number={1},
  pages={168--176},
  year={2002},
  publisher={IEEE}
}

@article{muzychka1999thermal,
  title={Thermal spreading resistance in multilayered contacts: applications in thermal contact resistance},
  author={Muzychka, YS and Sridhar, MR and Yovanovich, MM and Antonetti, VW},
  journal={Journal of thermophysics and heat transfer},
  volume={13},
  number={4},
  pages={489--494},
  year={1999}
}

@article{muzychka2004thermal,
  title={Thermal spreading resistance in compound and orthotropic systems},
  author={Muzychka, YS and Yovanovich, MM and Culham, JR},
  journal={Journal of thermophysics and heat transfer},
  volume={18},
  number={1},
  pages={45--51},
  year={2004}
}

@article{yovanovich2005four,
  title={Four decades of research on thermal contact, gap, and joint resistance in microelectronics},
  author={Yovanovich, M Michael},
  journal={IEEE transactions on components and packaging technologies},
  volume={28},
  number={2},
  pages={182--206},
  year={2005},
  publisher={IEEE}
}

@article{christofferson2008microscale,
  title={Microscale and nanoscale thermal characterization techniques},
  author={Christofferson, J and Maize, K and Ezzahri, Y and Shabani, J and Wang, X and Shakouri, A},
  year={2008}
}

@article{deshmukh2022direct,
  title={Direct measurement of nanoscale filamentary hot spots in resistive memory devices},
  author={Deshmukh, Sanchit and Rojo, Miguel Mu{\~n}oz and Yalon, Eilam and Vaziri, Sam and Koroglu, Cagil and Islam, Raisul and Iglesias, Ricardo A and Saraswat, Krishna and Pop, Eric},
  journal={Science advances},
  volume={8},
  number={13},
  pages={eabk1514},
  year={2022},
  publisher={American Association for the Advancement of Science}
}

@article{jain2016thermal,
  title={Thermal transport by phonons and electrons in aluminum, silver, and gold from first principles},
  author={Jain, Ankit and McGaughey, Alan JH},
  journal={Physical review B},
  volume={93},
  number={8},
  pages={081206},
  year={2016},
  publisher={APS}
}

@book{kittel2018introduction,
  title={Introduction to solid state physics},
  author={Kittel, Charles and McEuen, Paul},
  year={2018},
  publisher={John Wiley \& Sons}
}

@article{gu2002imaging,
  title={Imaging of thermal conductivity with sub-micrometer resolution using scanning thermal microscopy},
  author={Gu, YQ and Ruan, XL and Han, L and Zhu, DZ and Sun, XY},
  journal={International journal of thermophysics},
  volume={23},
  number={4},
  pages={1115--1124},
  year={2002},
  publisher={Springer}
}

@book{zhang2007nano,
  title={Nano/microscale heat transfer},
  author={Zhang, Zhuomin M and Zhang, Zhuomin M and Luby},
  volume={410},
  year={2007},
  publisher={Springer}
}

@article{tiwari2013anisotropic,
  title={Anisotropic thermal conductivity of thin polycrystalline oxide samples},
  author={Tiwari, Abhishek and Boussois, K{\'e}vin and Nait-Ali, Beno{\^\i}t and Smith, David Stanley and Blanchart, Philippe},
  journal={AIP Advances},
  volume={3},
  number={11},
  year={2013},
  publisher={AIP Publishing}
}

@article{zeng2017photothermal,
  title={Photothermal microscopy of coupled nanostructures and the impact of nanoscale heating in surface-enhanced Raman spectroscopy},
  author={Zeng, Zhi-Cong and Wang, Hao and Johns, Paul and Hartland, Gregory V and Schultz, Zachary D},
  journal={The Journal of Physical Chemistry C},
  volume={121},
  number={21},
  pages={11623--11631},
  year={2017},
  publisher={ACS Publications}
}

@article{isotta2023microscale,
  title={Microscale imaging of thermal conductivity suppression at grain boundaries},
  author={Isotta, Eleonora and Jiang, Shizhou and Moller, Gregory and Zevalkink, Alexandra and Snyder, G Jeffrey and Balogun, Oluwaseyi},
  journal={Advanced Materials},
  volume={35},
  number={38},
  pages={2302777},
  year={2023},
  publisher={Wiley Online Library}
}

@INPROCEEDINGS{10019538,
  author={Kelleher, A.B.},
  booktitle={2022 International Electron Devices Meeting (IEDM)}, 
  title={Celebrating 75 years of the transistor A look at the evolution of Moore's Law innovation}, 
  year={2022},
  volume={},
  number={},
  pages={1.1.1-1.1.5},
  doi={10.1109/IEDM45625.2022.10019538}}

@article{cahill2003nanoscale,
  title={Nanoscale thermal transport},
  author={Cahill, David G and Ford, Wayne K and Goodson, Kenneth E and Mahan, Gerald D and Majumdar, Arun and Maris, Humphrey J and Merlin, Roberto and Phillpot, Simon R},
  journal={Journal of applied physics},
  volume={93},
  number={2},
  pages={793--818},
  year={2003},
  publisher={American Institute of Physics}
}

@article{ostroverkhova2016organic,
  title={Organic optoelectronic materials: mechanisms and applications},
  author={Ostroverkhova, Oksana},
  journal={Chemical reviews},
  volume={116},
  number={22},
  pages={13279--13412},
  year={2016},
  publisher={ACS Publications}
}

@article{huang2020recent,
  title={Recent progress in organic phototransistors: Semiconductor materials, device structures and optoelectronic applications},
  author={Huang, Xianhui and Ji, Deyang and Fuchs, Harald and Hu, Wenping and Li, Tao},
  journal={ChemPhotoChem},
  volume={4},
  number={1},
  pages={9--38},
  year={2020},
  publisher={Wiley Online Library}
}

@article{munde20253d,
  title={3D integrated system for advanced intelligent computing},
  author={Munde, Ram Eknath and Vaillancourt, Noah and Chuang, Heng-Ray and Gu, Chongke and Wang, Yifan and Islam, Raisul},
  journal={Advances in Physics: X},
  volume={10},
  number={1},
  pages={2599301},
  year={2025},
  publisher={Taylor \& Francis}
}

@article{zhu2019development,
  title={Development trends and perspectives of future sensors and MEMS/NEMS},
  author={Zhu, Jianxiong and Liu, Xinmiao and Shi, Qiongfeng and He, Tianyiyi and Sun, Zhongda and Guo, Xinge and Liu, Weixin and Sulaiman, Othman Bin and Dong, Bowei and Lee, Chengkuo},
  journal={Micromachines},
  volume={11},
  number={1},
  pages={7},
  year={2019},
  publisher={MDPI}
}

@book{lyshevski2018mems,
  title={MEMS and NEMS: systems, devices, and structures},
  author={Lyshevski, Sergey Edward},
  year={2018},
  publisher={CRC press}
}

@article{putranto2024deep,
  title={A Deep Inside Quantum Technology Industry Trends and Future Implications},
  author={Putranto, Dedy Septono Catur and Wardhani, Rini Wisnu and Ji, Janghyun and Kim, Howon},
  journal={IEEE Access},
  year={2024},
  publisher={IEEE}
}

@article{hannah2020developments ,
  title={Developments in microscale and nanoscale sensors for biomedical sensing},
  author={Hannah, Stuart and Blair, Ewen and Corrigan, Damion K},
  journal={Current Opinion in Electrochemistry},
  volume={23},
  pages={7--15},
  year={2020},
  publisher={Elsevier}
}

@article{wang2013advances,
  title={Advances in nano-scaled biosensors for biomedical applications},
  author={Wang, Jianling and Chen, Guihua and Jiang, Hui and Li, Zhiyong and Wang, Xuemei},
  journal={Analyst},
  volume={138},
  number={16},
  pages={4427--4435},
  year={2013},
  publisher={Royal Society of Chemistry}
}

@article{tovar2023recent,
  title={Recent progress in micro-and nanotechnology-enabled sensors for biomedical and environmental challenges},
  author={Tovar-Lopez, Francisco J},
  journal={Sensors},
  volume={23},
  number={12},
  pages={5406},
  year={2023},
  publisher={MDPI}
}

@article{huang2019thermal,
  title={Thermal conductivity reduction in a silicon thin film with nanocones},
  author={Huang, Xin and Gluchko, Sergei and Anufriev, Roman and Volz, Sebastian and Nomura, Masahiro},
  journal={ACS applied materials \& interfaces},
  volume={11},
  number={37},
  pages={34394--34398},
  year={2019},
  publisher={ACS Publications}
}

@article{banerjee20023,
  title={3-D ICs: A novel chip design for improving deep-submicrometer interconnect performance and systems-on-chip integration},
  author={Banerjee, Kaustav and Souri, Shukri J and Kapur, Pawan and Saraswat, Krishna C},
  journal={Proceedings of the IEEE},
  volume={89},
  number={5},
  pages={602--633},
  year={2002},
  publisher={IEEE}
}

@article{pernot2021frequency,
  title={Frequency domain analysis of 3$\omega$-scanning thermal microscope probe---Application to tip/surface thermal interface measurements in vacuum environment},
  author={Pernot, Gilles and Metjari, A and Chaynes, H and Weber, M and Isaiev, Mykola and Lacroix, D},
  journal={Journal of Applied Physics},
  volume={129},
  number={5},
  year={2021},
  publisher={AIP Publishing}
}

@article{puyoo2010thermal,
  title={Thermal exchange radius measurement: Application to nanowire thermal imaging},
  author={Puyoo, Etienne and Grauby, St{\'e}phane and Rampnoux, Jean-Michel and Rouvi{\`e}re, Emmanuelle and Dilhaire, Stefan},
  journal={Review of Scientific Instruments},
  volume={81},
  number={7},
  year={2010},
  publisher={AIP Publishing}
}

@article{bahrami2004thermal,
  title={Thermal contact resistance of nonconforming rough surfaces, part 2: thermal model},
  author={Bahrami, MMYM and Culham, J Richard and Yovanovich, Michael M and Schneider, Gerry E},
  journal={Journal of Thermophysics and Heat Transfer},
  volume={18},
  number={2},
  pages={218--227},
  year={2004}
}

@article{swoboda2023spatially,
  title={Spatially-Resolved thermometry of filamentary nanoscale hot spots in TIO2 resistive random access memories to address device variability},
  author={Swoboda, Timm and Gao, Xing and Ros{\'a}rio, Carlos MM and Hui, Fei and Zhu, Kaichen and Yuan, Yue and Deshmukh, Sanchit and K{"o}ro{\u{g}}lu, {\c{C}}a{\u{g}}{\i}l and Pop, Eric and Lanza, Mario and others},
  journal={ACS Applied Electronic Materials},
  volume={5},
  number={9},
  pages={5025--5031},
  year={2023},
  publisher={ACS Publications}
}

@book{spiece2019quantitative,
  title={Quantitative mapping of nanothermal transport via Scanning Thermal Microscopy},
  author={Spi{\`e}ce, Jean},
  year={2019},
  publisher={Springer Nature}
}

@article{zhang2023realizing,
  title={Realizing the Accurate Measurements of Thermal Conductivity over a Wide Range by Scanning Thermal Microscopy Combined with Quantitative Prediction of Thermal Contact Resistance},
  author={Zhang, Qingqing and Zhu, Wei and Zhou, Jie and Deng, Yuan},
  journal={Small},
  volume={19},
  number={32},
  pages={2300968},
  year={2023},
  publisher={Wiley Online Library}
}

@article{yang2018liquid,
  title={Liquid-like, self-healing aluminum oxide during deformation at room temperature},
  author={Yang, Yang and Kushima, Akihiro and Han, Weizhong and Xin, Huolin and Li, Ju},
  journal={Nano letters},
  volume={18},
  number={4},
  pages={2492--2497},
  year={2018},
  publisher={ACS Publications}
}

@article{gorobez2021growth,
  title={Growth of Self-Passivating Oxide Layers on Aluminum---Pressure and Temperature Dependence},
  author={Gorobez, J{\"u}rgen and Maack, Bj{\"o}rn and Nilius, Niklas},
  journal={physica status solidi (b)},
  volume={258},
  number={5},
  pages={2000559},
  year={2021},
  publisher={Wiley Online Library}
}

@article{gomes2015scanning,
  title={Scanning thermal microscopy: A review},
  author={Gom{\`e}s, S{\'e}verine and Assy, Ali and Chapuis, Pierre-Olivier},
  journal={physica status solidi (a)},
  volume={212},
  number={3},
  pages={477--494},
  year={2015},
  publisher={Wiley Online Library}
}

@article{yovanovich2003thermal,
  title={Thermal Spreading and Contact},
  author={Yovanovich, MM and Marotta, EE},
  journal={Heat transfer handbook},
  volume={1},
  pages={261},
  year={2003},
  publisher={Wiley-Interscience}
}

@article{kennedy1960spreading,
  title={Spreading resistance in cylindrical semiconductor devices},
  author={Kennedy, David P},
  journal={Journal of applied physics},
  volume={31},
  number={8},
  pages={1490--1497},
  year={1960},
  publisher={American Institute of Physics}
}

@article{stoner1993kapitza,
  title={Kapitza conductance and heat flow between solids at temperatures from 50 to 300 K},
  author={Stoner, RJ and Maris, HJ},
  journal={Physical Review B},
  volume={48},
  number={22},
  pages={16373},
  year={1993},
  publisher={APS}
}

@article{hopkins2010criteria,
  title={Criteria for cross-plane dominated thermal transport in multilayer thin film systems during modulated laser heating},
  author={Hopkins, Patrick E and Serrano, Justin R and Phinney, Leslie M and Kearney, Sean P and Grasser, Thomas W and Harris, C Thomas},
  year={2010}
}

@article{cheaito2015thermal,
  title={Thermal boundary conductance accumulation and interfacial phonon transmission: Measurements and theory},
  author={Cheaito, Ramez and Gaskins, John T and Caplan, Matthew E and Donovan, Brian F and Foley, Brian M and Giri, Ashutosh and Duda, John C and Szwejkowski, Chester J and Constantin, Costel and Brown-Shaklee, Harlan J and others},
  journal={Physical Review B},
  volume={91},
  number={3},
  pages={035432},
  year={2015},
  publisher={APS}
}

@article{lombard2014influence,
  title={Influence of the electron--phonon interfacial conductance on the thermal transport at metal/dielectric interfaces},
  author={Lombard, J and Detcheverry, Fran{\c{c}}ois and Merabia, Samy},
  journal={Journal of Physics: Condensed Matter},
  volume={27},
  number={1},
  pages={015007},
  year={2014},
  publisher={IOP Publishing}
}

@article{schmidt2008pulse,
  title={Pulse accumulation, radial heat conduction, and anisotropic thermal conductivity in pump-probe transient thermoreflectance},
  author={Schmidt, Aaron J and Chen, Xiaoyuan and Chen, Gang},
  journal={Review of Scientific Instruments},
  volume={79},
  number={11},
  year={2008},
  publisher={AIP Publishing}
}

@article{kwon2021thermal,
  title={Thermal characterization of metal--oxide interfaces using time-domain thermoreflectance with nanograting transducers},
  author={Kwon, Heungdong and Perez, Christopher and Park, Woosung and Asheghi, Mehdi and Goodson, Kenneth E},
  journal={ACS Applied Materials \& Interfaces},
  volume={13},
  number={48},
  pages={58059--58065},
  year={2021},
  publisher={ACS Publications}
}

@inproceedings{fuchs1938conductivity,
  title={The conductivity of thin metallic films according to the electron theory of metals},
  author={Fuchs, K},
  booktitle={Mathematical Proceedings of the Cambridge Philosophical Society},
  volume={34},
  number={1},
  pages={100--108},
  year={1938},
  organization={Cambridge University Press}
}

@article{lugo2016thermal,
  title={Thermal properties of metallic films at room conditions by the heating slope},
  author={Lugo, JM and Oliva, AI},
  journal={Journal of Thermophysics and Heat Transfer},
  volume={30},
  number={2},
  pages={452--460},
  year={2016},
  publisher={American Institute of Aeronautics and Astronautics}
}

@article{mayadas1970electrical,
  title={Electrical-resistivity model for polycrystalline films: the case of arbitrary reflection at external surfaces},
  author={Mayadas, AF and Shatzkes, M},
  journal={Physical review B},
  volume={1},
  number={4},
  pages={1382},
  year={1970},
  publisher={APS}
}

@article{zhu2010ultrafast,
  title={Ultrafast thermoreflectance techniques for measuring thermal conductivity and interface thermal conductance of thin films},
  author={Zhu, Jie and Tang, Dawei and Wang, Wei and Liu, Jun and Holub, Kristopher W and Yang, Ronggui},
  journal={Journal of Applied Physics},
  volume={108},
  number={9},
  year={2010},
  publisher={AIP Publishing}
}

@article{Volkov1976AluminumThermal,
  author    = {Volkov, L. S. and Palatnik, L. S. and Pugachev, A. T.},
  title     = {Investigation of the Thermal Properties of Thin Aluminum Films},
  journal   = {Soviet Physics JETP},
  volume    = {43},
  number    = {6},
  pages     = {1171--1174},
  year      = {1976},
  note      = {Translated from Zhurnal Eksperimental'noi i Teoreticheskoi Fiziki, Vol. 70, p. 2244 (1976)},
  url       = {https://www.jetp.ras.ru/cgi-bin/dn/e_043_06_1171.pdf}
}

@article{Wu1993PhotothermalThinFilms,
  author    = {Wu, Z. and P. Kuo and W. Lanhua and S. Gu and R. Thomas},
  title     = {Photothermal characterization of optical thin films},
  journal   = {Thin Solid Films},
  volume    = {236},
  number    = {1},
  pages     = {191--198},
  year      = {1993},
  doi       = {10.1016/0040-6090(93)90668-F},
  url       = {https://www.sciencedirect.com/science/article/pii/004060909390668F}
}

@article{Jin2008ElectronMeanFreePath,
  author    = {Jin, Jae Sik and Lee, Joon Sik and Kwon, Ohmyoung},
  title     = {Electron effective mean free path and thermal conductivity predictions of metallic thin films},
  journal   = {Applied Physics Letters},
  volume    = {92},
  number    = {17},
  pages     = {171910},
  year      = {2008},
  doi       = {10.1063/1.2917454},
  url       = {https://doi.org/10.1063/1.2917454}
}

@article{Schmidt2010ThinMetalFDTR,
  author    = {Schmidt, Aaron J. and Cheaito, Ramez and Chiesa, Matteo},
  title     = {Characterization of thin metal films via frequency-domain thermoreflectance},
  journal   = {Journal of Applied Physics},
  volume    = {107},
  number    = {2},
  pages     = {024908},
  year      = {2010},
  doi       = {10.1063/1.3289907},
  url       = {https://doi.org/10.1063/1.3289907}
}

@article{Sondheimer1952MeanFreePath,
  author    = {Sondheimer, E. H.},
  title     = {The mean free path of electrons in metals},
  journal   = {Advances in Physics},
  volume    = {1},
  number    = {1},
  pages     = {1--42},
  year      = {1952},
  doi       = {10.1080/00018735200101151},
  url       = {https://doi.org/10.1080/00018735200101151}
}

@article{Gall2016,
  author  = {Gall, Daniel},
  title   = {Electron mean free path in elemental metals},
  journal = {Journal of Applied Physics},
  volume  = {119},
  number  = {8},
  pages   = {085101},
  year    = {2016},
  doi     = {10.1063/1.4942216}
}

\end{document}